\documentclass[doublecol]{epl2} 

\usepackage{graphicx,amsfonts,amssymb,amsmath}
\usepackage{textcomp}
\usepackage{esint}
\usepackage{lmodern}
\usepackage[dvipsnames]{xcolor}
\usepackage[english]{babel}
\usepackage{afterpage}
\usepackage{bbold}

\title{Is there a Mott-glass phase in a one-dimensional disordered quantum fluid with linearly confining interactions?}
\shorttitle{Disordered one-dimensional bosons with linearly confining interactions} 

\author{Nicolas Dupuis}

\institute{Sorbonne Universit\'e, CNRS, Laboratoire de Physique Th\'eorique de la Mati\`ere Condens\'ee, LPTMC, \\ F-75005 Paris, France 
}

\abstract{We study a one-dimensional disordered quantum fluid with linearly confining interactions (disordered Schwinger model) using bosonization and the nonperturbative functional renormalization group. We find that the long-range interactions make the Anderson-insulator (or, for bosons, the Bose-glass) fixed point (corresponding to a compressible state with a gapless optical conductivity) unstable, even if the latter may control the flow at intermediate energy scales. The stable fixed point describes an incompressible ground state with a gapped optical conductivity similar to a Mott insulator. These results disagree with the Gaussian variational method that predicts a Mott glass, namely a state with vanishing compressibility but a gapless optical conductivity.}

\begin{document}


\def\rhoeq{\hat\rho_{\rm eq}}

\newcommand{\marge}[1]{}
\newcommand{\remarque}[1]{}
\newcommand{\new}[1]{{\bf #1}}
\newlength{\textlarg}
\newcommand{\redbar}[1]{\textcolor{red}{\st{#1}}} 
\newcommand{\bluebar}[1]{\textcolor{blue}{\st{#1}}} 

\newcommand{\beq}{\begin{equation}}
\newcommand{\eeq}{\end{equation}}
\newcommand{\bfig}{\begin{figure}}
\newcommand{\efig}{\end{figure}}
\newcommand{\bline}{\begin{multline}}
\newcommand{\eline}{\end{multline}}
\newcommand{\bremark}{\begin{quotation} \noindent \small }
\newcommand{\eremark}{\end{quotation}}
\newcommand{\llbrace}{\left\lbrace}  
\newcommand{\rrbrace}{\right\rbrace}
\newcommand{\lbraket}{\left[}
\newcommand{\rbraket}{\right]}
\newcommand{\llangle}{\left\langle}
\newcommand{\rrangle}{\right\rangle} 

\newcommand{\Tr}{{\rm Tr}} 
\newcommand{\tr}{{\rm tr}} 
\newcommand{\sgn}{\,{\rm sgn}} 
\newcommand{\mean}[1]{\langle #1 \rangle}
\newcommand{\commu}[2]{[#1,#2]} 
\newcommand{\bra}[1]{\langle#1|}
\newcommand{\ket}[1]{|#1\rangle}
\newcommand{\braket}[2]{\langle #1|#2\rangle}
\newcommand{\ketbra}[2]{|#1\rangle\langle#2|}
\newcommand{\dbraket}[3]{\langle #1|#2|#3\rangle}
\newcommand{\vac}{|{\rm vac}\rangle} 
\newcommand{\bravac}{\langle{\rm vac}|}
\newcommand{\const}{{\rm const}} 
\newcommand{\atanh}{\,{\rm atanh}}
\newcommand{\cotanh}{\,{\rm cotanh}}

\newcommand{\ie}{i.e.\xspace}
\newcommand{\iet}{i.e.}
\newcommand{\eg}{e.g.\xspace}
\newcommand{\cc}{{\rm c.c.}} 
\newcommand{\hc}{{\rm h.c.}} 
\newcommand\eme{$^{\mbox{\footnotesize ème}}$\xspace}

\newcommand{\jhatbf}{\hat {\textbf \jold}} 
\newcommand{\Jhatbf}{\hat {\textbf \J}} 
\newcommand{\jhat}{\hat {\jmath}} 
\newcommand{\Jhat}{\hat {J}} 
\newcommand{\jbf}{\textbf j}
\newcommand{\Jbf}{\textbf J}

\def\chibf{\boldsymbol{\chi}}
\def\down{\downarrow}
\def\eps{\epsilon}
\def\gam{\gamma} 
\def\alphabf{\boldsymbol{\alpha}}
\def\phibf{\boldsymbol{\phi}}
\def\varphibf{\boldsymbol{\varphi}}
\def\varphibfs{\boldsymbol{\varphi}_<}
\def\varphibfl{\boldsymbol{\varphi}_>}
\def\varphis{\varphi_{<}}
\def\varphil{\varphi_{>}}
\def\psibf{\boldsymbol{\psi}}
\def\thetabf{\boldsymbol{\theta}}
\def\Ome{\Omega}
\def\omeD{{\omega_D}} 
\def\bfOme{\boldsymbol{\Omega}} 
\def\Omebf{\boldsymbol{\Omega}} 
\def\lamb{\lambda}
\def\Lamb{\Lambda}
\def\sig{\sigma}
\def\Sig{\Sigma}
\def\sigp{{\sigma'}} 
\def\bfsig{\boldsymbol{\sigma}} 
\def\sigbf{\boldsymbol{\sigma}} 
\def\bfSig{\boldsymbol{\Sigma}} 
\def\The{\Theta} 
\def\up{\uparrow}

\def\epsk{\epsilon_{\bf k}} 
\def\xik{\xi_{\bf k}} 
\def\txik{\tilde\xi_{\bf k}} 
\def\xip{\xi_{\bf p}} 
\def\xiq{\xi_{\bf q}} 
\def\xikq{\xi_{{\bf k}+{\bf q}}} 
\def\Ek{E_{\bf k}} 
\def\Ep{E_{\bf p}}
\def\Eq{E_{\bf q}}
\def\Heff{\hat H_{\rm eff}}
\def\Hem{\hat H_{\rm em}}
\def\Hint{\hat H_{\rm int}}
\def\Hloc{\hat H_{\rm loc}}
\def\HMF{\hat H_{\rm MF}}
\def\Sem{S_{\rm em}}
\def\SMF{S_{\rm MF}} 
\def\SHF{S_{\rm HF}} 
\def\SRPA{S_{\rm RPA}} 
\def\Sint{S_{\rm int}} 
\def\Sloc{S_{\rm loc}}
\def\TN{T_{\rm N}} 
\def\TNHF{T^{\rm HF}_{\rm N}} 
\def\Zloc{Z_{\rm loc}} 
\def\ZMF{Z_{\rm MF}} 
\def\ZHF{Z_{\rm HF}} 
\def\ZRPA{Z_{\rm RPA}} 
\def\RPA{{\rm RPA}}
\def\loc{{\rm loc}} 
\def\pp{{\rm pp}}
\def\ph{{\rm ph}} 
\def\ch{{\rm ch}}
\def\sp{{\rm sp}} 
\def\qtf{q_{\rm TF}}
\def\epstf{\eps^{}_{\rm TF}} 
\def\epsrpa{\eps^{}_{\rm RPA}} 
\def\chinnzpp{\chi_{nn}^{0}{}\!\!\!''}

\def\half{\frac{1}{2}}
\def\dhalf{\dfrac{1}{2}}
\def\third{\frac{1}{3}} 
\def\quarter{\frac{1}{4}}

\def\qr{{\bf q}\cdot{\bf r}}
\def\wt{\omega t} 

\def\a{{\bf a}}
\def\b{{\bf b}}
\newcommand{\cv}{{\bf c}} 
\def\e{{\bf e}}
\def\f{{\bf f}}
\def\g{{\bf g}}
\def\h{{\bf h}}
\def\jold{\char"11}
\def\j{{\bf j}}
\def\k{{\bf k}}
\def\l{{\bf l}}
\def\m{{\bf m}}
\def\n{{\bf n}} 
\def\p{{\bf p}} 
\def\q{{\bf q}}
\def\r{{\bf r}}
\def\t{{\bf t}}
\def\u{{\bf u}}
\newcommand{\vv}{{\bf v}}
\def\x{{\bf x}}
\def\y{{\bf y}} 
\def\z{{\bf z}} 
\def\A{{\bf A}}
\def\B{{\bf B}}
\def\D{{\bf D}} 
\def\E{{\bf E}} 
\def\F{{\bf F}} 
\def\H{{\bf H}}  
\def\J{{\bf J}}
\def\K{{\bf K}} 

\def\G{{\bf G}}
\def\L{{\bf L}}
\def\M{{\bf M}}  
\def\O{{\bf O}} 
\def\P{{\bf P}} 
\def\Q{{\bf Q}} 
\def\R{{\bf R}}
\def\S{{\bf S}}
\def\U{{\bf U}} 
\def\V{{\bf V}} 
\def\X{{\bf X}} 
\def\Y{{\bf Y}} 
\def\epsbf{\boldsymbol{\epsilon}}
\def\betabf{\boldsymbol{\beta}}
\def\deltabf{\boldsymbol{\delta}}
\def\mubf{\boldsymbol{\mu}}
\def\nablabf{\boldsymbol{\nabla}}
\def\rhobf{\boldsymbol{\rho}}
\def\sigmabf{\boldsymbol{\sigma}} 
\def\Pibf{\boldsymbol{\Pi}}
\def\pibf{\boldsymbol{\pi}}

\def\para{\parallel}
\def\kpara{{k_\parallel}}
\def\kperp{{k_\perp}} 
\def\kperpp{{k_\perp'}} 
\def\qperp{{q_\perp}} 
\def\tperp{{t_\perp}} 

\def\w{\omega}
\def\wn{\omega_n}
\def\wm{\omega_m}
\def\wnu{\omega_\nu}
\def\wp{\omega_p} 
\def\dmu{{\partial_\mu}}
\def\dnu{{\partial_\nu}}
\def\dl{{\partial_l}}  
\def\dt{\partial_t} 
\def\tdt{\tilde\partial_t}
\def\dk{\partial_k}
\def\tdk{\tilde\partial_k}
\def\dx{\partial_x}
\def\dy{\partial_y} 
\def\dtau{{\partial_\tau}}  
\def\det{{\rm det}} 
\def\Pf{{\rm Pf}}
\def\diag{{\rm diag}}

\def\dsum{\displaystyle \sum}
\def\dint{\displaystyle \int} 
\def\intt{\int_{-\infty}^\infty dt} 
\def\inttp{\int_{-\infty}^\infty dt'} 
\def\intk{\int_{\bf k}} 
\def\intkd{\int \frac{d^dk}{(2\pi)^d}}
\def\intq{\int_{\bf q}} 
\def\intr{\int d^dr}  
\def\dintr{\displaystyle \int d^dr} 
\def\intrp{\int d^dr'}
\def\dinttau{\displaystyle \int_0^\beta d\tau}
\def\dinttaup{\displaystyle \int_0^\beta d\tau'}
\def\inttau{\int_0^\beta d\tau}
\def\inttaup{\int_0^\beta d\tau'}
\def\intx{\int d^{d+1}x} 
\def\inttaur{\int_0^\beta d\tau \int d^dr}
\def\intinf{\int_{-\infty}^\infty}
\def\dinttaur{\displaystyle \int_0^\beta d\tau \int d^dr}
\def\dintinf{\displaystyle \int_{-\infty}^\infty}
\def\intw{\int_{-\infty}^\infty \frac{d\w}{2\pi}}
\def\sumr{\sum_{\bf r}} 

\def\calA{{\cal A}}
\def\calAbf{\bm{{\cal A}}}
\def\calB{{\cal B}} 
\def\calC{{\cal C}} 
\def\dt{\partial_t}
\def\calD{{\cal D}}
\def\calE{{\cal E}}
\def\calF{{\cal F}} 
\def\calFbf{\bm{{\cal F}}}
\def\calG{{\cal G}}
\def\calH{{\cal H}}
\def\calI{{\cal I}}
\def\calJ{{\cal J}}
\def\calK{{\cal K}}
\def\calL{{\cal L}} 
\def\calM{{\cal M}} 
\def\calN{{\cal N}}
\def\calO{{\cal O}}
\def\calP{{\cal P}}  
\def\calR{{\cal R}} 
\def\calS{{\cal S}}
\def\calT{{\cal T}}
\def\calU{{\cal U}}
\def\calV{{\cal V}}
\def\calX{{\cal X}} 
\def\calY{{\cal Y}} 
\def\calZ{{\cal Z}} 

\def\calbfB{{\bf \cal B}}
\def\calbfF{{\bf \cal F}}

\def\tT{{\tilde T}}
\def\talpha{{\tilde\alpha}}
\def\tbeta{{\tilde\beta}}
\def\tchi{{\tilde\chi}}
\def\tdelta{{\tilde\delta}}
\def\tDelta{{\tilde\Delta}}
\def\teta{{\tilde\eta}} 
\def\tlamb{{\tilde\lambda}}
\def\tmu{{\tilde\mu}}
\def\tphibf{{\tilde\phibf}}
\def\trho{{\tilde\rho}}
\def\tvarphibf{{\tilde\varphibf}} 
\def\tw{{\tilde\omega}}
\def\twn{{\tilde\omega_n}}
\def\twnu{{\tilde\omega_\nu}}

\def\asinh{{\rm asinh}} 

\maketitle

\section{Introduction}

In quantum many-body systems the interplay between disorder and interactions may lead to rich phase diagrams exhibiting various types of insulating phases. In a one-dimensional quantum fluid, depending on the respective strength of disorder and interactions, the ground state can be either a Luttinger liquid (i.e a superfluid/metallic state for bosons/fermions) or an Anderson insulator (a Bose glass for bosons)~\cite{Giamarchi87,Giamarchi88}. The latter is characterized by a nonzero compressibility, a vanishing dc conductivity and the absence of a gap in the optical conductivity.   

An interesting question is whether one-dimensional quantum fluids can exhibit other, possibly more exotic, phases besides the Anderson-insulator/Bose-glass and Luttinger-liquid phases. It was suggested that in a Fermi fluid the interplay between disorder and a commensurate periodic potential could stabilize a Mott glass, characterized by a vanishing compressibility and a gapless optical conductivity~\cite{Orignac99,Giamarchi01}. This state, intermediate between a Mott insulator and an Anderson insulator, would result from the coexistence of gapped single-particle excitations (which imply a vanishing compressibility) and gapless particle-hole excitations (hence the absence of a gap in the conductivity). The realization of a Mott glass in a one-dimensional quantum fluid with short-range interactions has however been challenged~\cite{Nattermann07,Ledoussal08a}, thus calling into question the validity of the perturbative functional renormalization group (FRG) approach and the Gaussian variational method (GVM)~\cite{Mezard91,Giamarchi96} used in Refs.~\cite{Orignac99,Giamarchi01}.

In a recent paper~\cite{Chou18}, Chou {\it et al.} have studied a model of one-dimensional fermions interacting with a $(1+1)D$ gauge field (Schwinger model~\cite{Schwinger62,Coleman76,Fischler79,Wolf85}) in the presence of quenched disorder (disordered Schwinger model~\cite{Nandkishore17,Akhtar18}). This model evades the arguments against the Mott-glass phase advanced in Ref.~\cite{Nattermann07} due to the long-range nature of the linearly confining interactions mediated by the gauge field. Using bosonization and the Gaussian variational method (GVM), these authors find that the ground state is a Mott glass. This conclusion agrees with the result obtained in Ref.~\cite{Giamarchi01} within the perturbative FRG approach.\footnote{The perturbative FRG approach to the disordered Schwinger model is discussed in
	Appendix D of Ref.~\cite{Giamarchi01}.} The Schwinger model has experimental relevance since it has been proposed~\cite{Rico14,Kuhn14,Notarnicola15,Yang16} and realized~\cite{Martinez16} in synthetic quantum systems.

Given the uncontrolled nature of the GVM and the perturbative FRG (in phases where disorder flows to strong coupling) as well as their controversial results in models with both disorder and a commensurate periodic potential~\cite{Orignac99,Giamarchi01,Nattermann07,Ledoussal08a}, it would be desirable to confirm the existence of a Mott-glass phase in the disordered Schwinger model by alternative methods. In this Letter, we reconsider this issue using bosonization, the replica method, and the nonperturbative FRG. For simplicity, we consider bosons but our conclusions also hold for a Fermi fluid. In the absence of linearly confining interactions, i.e. for a disordered fluid with short-range interactions, the main features of the FRG flow are well understood~\cite{Dupuis19,Dupuis20}. There is an attractive line of fixed points at vanishing disorder corresponding to the superfluid phase. This line becomes repulsive when the Luttinger parameter $K$, which characterizes the quantum fluctuations of the particle density, becomes smaller than 3/2. All RG trajectories that do not end up in the superfluid phase are attracted by a fixed point characterized by a vanishing Luttinger parameter $K^*_{\rm BG}=0$ and a singular, cuspy, functional disorder correlator that signals the existence of metastable states and the glassy properties of the Bose glass.

In the bosonization formalism, the linearly confining interactions translate into a mere ``mass'' term for the phase field $\varphi$ describing density fluctuations. This mass term is a relevant perturbation that makes both the attractive line of fixed points corresponding to the superfluid Luttinger liquid and the Bose-glass fixed point unstable, whereas the new, stable, fixed point describes an incompressible ground state with a gapped optical conductivity similar to a Mott insulator. Nevertheless, when the strength of the linearly confining interactions is sufficiently weak, the flow is controlled by the Bose-glass fixed point at intermediate energy scales before crossing over to the Mott-insulator-like regime at lower energies. The possible reasons for the discrepancy between the nonperturbative FRG approach on the one hand, and the GVM and perturbative FRG approach on the other hand, will be discussed in the conclusion.

\section{Disordered Schwinger model} 

We consider a system of particles interacting with a dynamical gauge field in one spatial and one temporal dimension (Schwinger model~\cite{Schwinger62,Coleman76,Fischler79,Wolf85}). The gauge field is equivalent to a one-dimensional Coulomb potential $e^2/q^2$ and induces a linearly confining inter-particle potential $-(e^2/2)|x|$ where $e$ is the charge of the particles. The Hamiltonian of the system reads 
\beq 
\hat H_0 = \hat H_{\rm LL} - \frac{e^2}{4} \int_{x,x'} \hat\rho(x) |x-x'| \hat\rho(x') 
\eeq 
(we set $\hbar=k_B=1$ throughout the paper), where $\hat\rho$ is the density operator and the Luttinger-liquid Hamiltonian $\hat H_{\rm LL}$ includes the kinetic energy of the bosons and a short-range interaction. Using bosonization~\cite{Giamarchi_book}, at low energies the Hamiltonian can be written as 
\beq 
\hat H_0 = \int_x  \llbrace \frac{v}{2\pi} \left[ \frac{1}{K} (\dx\hat\varphi)^2 + K (\dx\hat\theta)^2 \right] 
+ \frac{e^2}{2\pi^2} \hat\varphi^2 \rrbrace , 
\eeq 
where $\hat\theta$ is the phase of the boson operator $\hat\psi(x)=e^{i\hat\theta(x)} \hat\rho(x)^{1/2}$ and $\hat\varphi$ is related to the density operator {\it via} $\hat\rho=\rho_0-\frac{1}{\pi}\dx\hat\varphi +2 \rho_2\cos(2\pi\rho_0x - 2\hat\varphi)$. Here $\rho_0$ is the average density of particles and $\rho_2$ a nonuniversal parameter that depends on microscopic details. $v$ denotes the sound mode velocity of the Luttinger liquid and the dimensionless quantity $K$, which encodes the strength of the short-range boson interactions, is the Luttinger parameter. In the bosonization formalism, the linearly confining interactions of the Schwinger model reduces to a mere quadratic ``mass'' term. 

On the other hand the disorder contributes to the Hamiltonian a term $\hat H_{\rm dis} =\int_x V(x)\hat\rho(x)$ where $V(x)$ is a random potential assumed to have a Gaussian probability distribution with zero mean. Ignoring the long-wavelength part of $V(x)$, which does not play any role in the localization~\cite{Chou18}, 
\beq
\hat H_{\rm dis} = \int_x \rho_2 ( \xi^* e^{2i\hat\varphi} + \hc ) ,
\eeq 
where the random (complex) potential $\xi(x)$ satisfies $\overline{\xi(x)}=0$ and  $\overline{\xi^*(x)\xi(x')}=(\calD/\rho_2^2)\delta(x-x')$ (the overbar denotes the average over disorder). The Hamiltonian $\hat H_0+\hat H_{\rm dis}$ is conveniently studied by considering $n$ copies (or replicas) of the system. After averaging over disorder, we can write the partition function as a functional integral over the density field $\varphi$ with the replicated action 
\begin{align}
S[\varphi] ={}& \sum_a \int_{x,\tau} \llbrace \frac{v}{2\pi K} \left[ (\dx\varphi_a)^2 + \frac{(\dtau\varphi_a)^2}{v^2} \right] + \frac{e^2}{2\pi^2} \varphi^2_a \rrbrace \nonumber \\ &
 -\calD \sum_{a,b} \int_{x,\tau,\tau'} \cos[2\varphi_a(x,\tau)-2\varphi_b(x,\tau')] , 
\label{action} 
\end{align}
where $\varphi=\{\varphi_a\}$ and $a,b=1\cdots n$ are replica indices. The imaginary times $\tau,\tau'$ vary in $[0,\beta]$ with $\beta=1/T\to\infty$. 

\section{FRG formalism}
In the nonperturbative FRG approach, one considers a family of models indexed by a momentum scale $k$ such that fluctuations
are smoothly taken into account as $k$ is lowered from the microscopic scale $\Lambda$ (the UV cutoff of the model) down to 0~\cite{Berges02,Delamotte12,Kopietz_book}. This is achieved by adding to the action~(\ref{action}) the infrared regulator term
\begin{equation}
\Delta S_k[\varphi] = \half \sum_{a,q,\w} \varphi_a(-q,-i\w) R_{k}(q,i\w) \varphi_a(q,i\w) ,
\label{DeltaSk} 
\end{equation}
where $\w\equiv\wm=2\pi m/\beta$ ($m$ integer) is a Matsubara frequency. The cutoff function $R_k(q,i\w)$ is chosen so that fluctuation modes satisfying $|q|,|\w|/v_k\ll k$ are suppressed while those with $|q|\gg k$ or $|\w|/v_k\gg k$ are left unaffected ($v_k$ denotes the $k$-dependent sound-mode velocity, see below); its precise form is given in Ref.~\cite{Dupuis20}. The partition function 
\beq 
\calZ_k[J] = \int \calD[\varphi] \, 
e^{-S[\varphi] - \Delta S_k[\varphi]+ \sum_a \int_{x,\tau} J_a\varphi_a } ,
\eeq
defined here in the presence of $n$ external sources $J=\{J_a\}$ acting on each replica independently, thus becomes $k$ dependent. The main quantity of interest in the FRG approach is the scale-dependent effective action 
\beq 
\Gamma_k[\phi] = - \ln \calZ_k[J] + \sum_a \int_{x,\tau} J_a \phi_a - \Delta S_k[\phi] , 
\eeq 
defined as a (slightly) modified Legendre transform that includes the subtraction of $\Delta S_k[\phi]$. Here $\phi=\{\phi_a\}$ and $\phi_a=\mean{\varphi_a}$ denotes the expectation value of $\varphi_a$. Assuming that for $k=\Lambda$ the fluctuations are completely frozen by the $\Delta S_\Lambda$ term, $\Gamma_\Lambda[\phi]=S[\phi]$. On the other hand the effective action of the original model~(\ref{action}) is given by $\Gamma_{k=0}$ provided that $R_{k=0}$ vanishes. The nonperturbative FRG aims at determining $\Gamma_{k=0}$ from $\Gamma_\Lambda$ using Wetterich's equation~\cite{Wetterich93,Ellwanger94,Morris94}
\beq
\dt \Gamma_k[\phi] = \half \Tr \left\{ \dt R_k \bigl(\Gamma_k^{(2)}[\phi] + R_k \bigr)^{-1} \right\} ,
\label{eqwet}
\eeq
where $\Gamma_k^{(2)}$ is the second functional derivative of $\Gamma_k$ and $t=\ln(k/\Lamb)$ a (negative) RG ``time''. The trace in~(\ref{eqwet}) involves a sum over momenta and frequencies as well as the replica index. 

The form of the effective action is strongly constrained by the statistical tilt symmetry (STS)~\cite{Schulz88} which originates from the invariance of the disorder part of the action~(\ref{action}) in the change $\varphi_a(x,\tau)\to \varphi_a(x,\tau)+w(x)$ with $w(x)$ an arbitrary time-independent function~\cite{Dupuis20}:  
\begin{align}
\Gamma_k[\phi] ={}& \Gamma_k[\phi'] - \frac{n}{2} \beta \int_x  \Bigl[ Z_x(\dx w)^2 + \frac{e^2}{\pi^2} w^2 \Bigr] \nonumber \\ &
 -  \int_{x,\tau} \sum_a  \Bigl[ Z_x(\dx w)(\dx \phi_a) + \frac{e^2}{\pi^2} \phi_a w \Bigr],  
\end{align} 
where $\phi'_a(x,\tau)=\phi_a(x,\tau)+w(x)$ and $Z_x=v/\pi K$. A possible ansatz for the effective action, which allows us to solve (approximately) the flow equation~(\ref{eqwet}) while being compatible with the STS, is given by  
\beq 
\Gamma_{k}[\phi] = \sum_a \Gamma_{1,k}[\phi_a] - \half \sum_{a,b} \Gamma_{2,k}[\phi_a,\phi_b] ,
\label{ansatz1} 
\eeq
where
\begin{align}
&\Gamma_{1,k}[\phi_a] = \int_{x,\tau} \llbrace \frac{Z_{x}}{2} (\dx\phi_a)^2 + \half \phi_a \Delta_k(-\dtau) \phi_a + \frac{e^2}{2\pi^2} \phi_a^2 \rrbrace \nonumber \\  
&\Gamma_{2,k}[\phi_a,\phi_b] = \int_{x,\tau\,\tau'} V_k(\phi_a(x,\tau)-\phi_b(x,\tau')) ,
\label{ansatz2}
\end{align}
with the initial conditions $\Delta_\Lambda(i\w)=\w^2/\pi vK$ and $V_\Lambda(u)=2\calD\cos(2u)$. The STS implies that the ``self-energy'' $\Delta_k(i\w)$ satisfies $\Delta_k(i\w=0)=0$, $Z_x$ remains equal to its initial value and no higher-order space derivatives are allowed. The infrared regulator $\Delta S_k$ ensures that $\Delta_k(i\w)=Z_x\w^2/v_k^2+\calO(\w^4)$ is a regular function near $\w=0$. In addition to the running velocity $v_k$ one may define a $k$-dependent Luttinger parameter by $Z_x=v_k/\pi K_k$. 

In the absence of linearly confining interactions ($e=0$) the ansatz~(\ref{ansatz1}) corresponds to the one used in Refs.~\cite{Dupuis19,Dupuis20} to study the Bose-glass phase. In the disordered Schwinger model the effective action includes the additional mass term $(e^2/2\pi^2) \phi_a^2$. The fact that this term is $k$ independent, as required by the STS, has a very important consequence: The propagator $P_k=1/\Gamma_{1,k}^{(2)}$ obtained from the one-replica term $\Gamma_{1,k}$ is massive, 
\beq
P_k(q,i\w) = \frac{1}{Z_xq^2+\Delta_k(i\w)+e^2/\pi^2} ,
\label{Pk} 
\eeq
and one can already anticipate that neither the Luttinger liquid nor the Bose glass can be realized. 

By inserting the ansatz~(\ref{ansatz1}-\ref{ansatz2}) into Eq.~(\ref{eqwet}), we obtain the flow equations
\beq
\begin{split} 
\dt\delta_k(u) ={}& -3 \delta_k(u) - K_k l_1 \delta''_k(u) \\ &
+ \pi \bar l_2 [ \delta_k''(u) (\delta_k(u)-\delta_k(0)) + \delta'_k(u)^2 ] ,  \\ 
\dt \tilde\Delta_k(i\tw) ={}& - 2 \tilde\Delta_k(i\tw) + z_k \tw \partial_{\tw} \tilde\Delta_k(i\tw)  \\ &
- \pi \delta''_k(0) [ \bar l_1(i\tw) - \bar l_1(0) ] ,  \\ 
\dt K_k ={}& \theta_k K_k , \qquad
\dt (K_k/v_k) = 0 ,  
\end{split}
\label{rgeq}
\eeq
for the dimensionless functions 
\begin{equation}
\delta_k(u) = - \frac{K^2}{v^2} \frac{V_k''(u)}{k^3} , \quad  
\tilde \Delta_k(i\tw) = \frac{\Delta(i\w)}{Z_x k^2}  \label{dimvar} 
\end{equation}
$(\tw=\w/v_kk$), where $z_k=1+\theta_k$ is the running dynamical critical exponent and  
\begin{equation}
\theta_k = \frac{\pi}{2} \delta''_k(0) \bar m_\tau . 
\label{thetak}
\end{equation}
These equations are similar to those derived in Ref.~\cite{Dupuis20}. The only modification coming from the linearly confining interaction is the presence of the nonzero mass in the propagator~(\ref{Pk}) which enters the ``threshold'' functions $l_1,\bar l_2,\bar l_1(i\tw),\bar m_\tau$ (see Appendix D4 in Ref.~\cite{Dupuis20} for an explicit definition of the threshold functions). 

Note that the propagator in dimensionless form, $\tilde P_k(\tilde q,i\tw)=Z_xk^2 P_k(q,i\w)$ with $\tilde q=q/k$, is naturally expressed in terms of the dimensionless square charge $\tilde e^2_k=e^2/Z_xk^2$. The latter increases as $k\to 0$ and is therefore a relevant perturbation. Thus one can already anticipate the existence of two regimes in the RG flow: a short-distance or high-momentum regime, defined by $\tilde e^2_k/\pi^2\ll 1$ (or, equivalently, $k\gg k_x$ where $k_x=e\sqrt{K/\pi v}$), where the physics is dominated by the quenched disorder and the short-range interactions, and a long-distance regime $\tilde e^2_k/\pi^2\gg 1$ where the linearly confining interactions play an essential role. We expect that in the latter regime, the mass term acts as an effective infrared cutoff and stops the RG flow.

\section{Ground state of the disordered Schwinger model} 

Let us first briefly recall the main results of the FRG analysis when $e^2=0$~\cite{Dupuis19,Dupuis20}. Besides the attractive line of fixed points defined by $K\geq 3/2$ and $\delta(u)=0$, there is a fixed point, describing the Bose-glass phase, characterized by a vanishing Luttinger parameter $K^*_{\rm BG}=0$ and a potential $\delta^*_{\rm BG}(u)$ exhibiting cusps at $u=0,\pm\pi,\pm 2\pi$, etc. At finite momentum scales $k$, the cusp singularity is rounded into a quantum boundary layer (QBL) with a width $\sim K_k\sim k^{z-1}$ determined by the dynamical critical exponent $z=\lim_{k\to 0}z_k$. The QBL controls the low-energy dynamics of the Bose-glass phase and in particular yields a (dissipative) conductivity vanishing as $\w^2$ in the low-frequency limit. 

\begin{figure}
\centerline{\includegraphics[width=4.1cm]{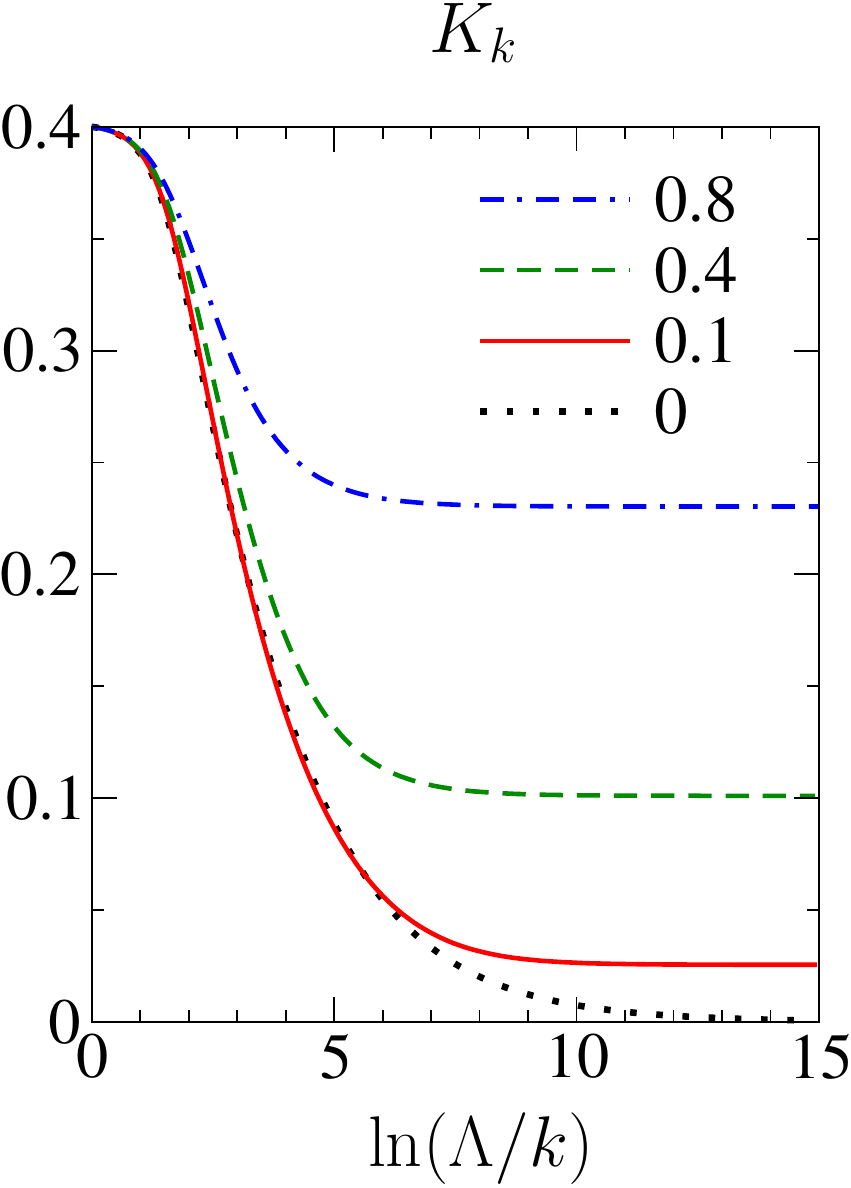}
\includegraphics[width=4.03cm]{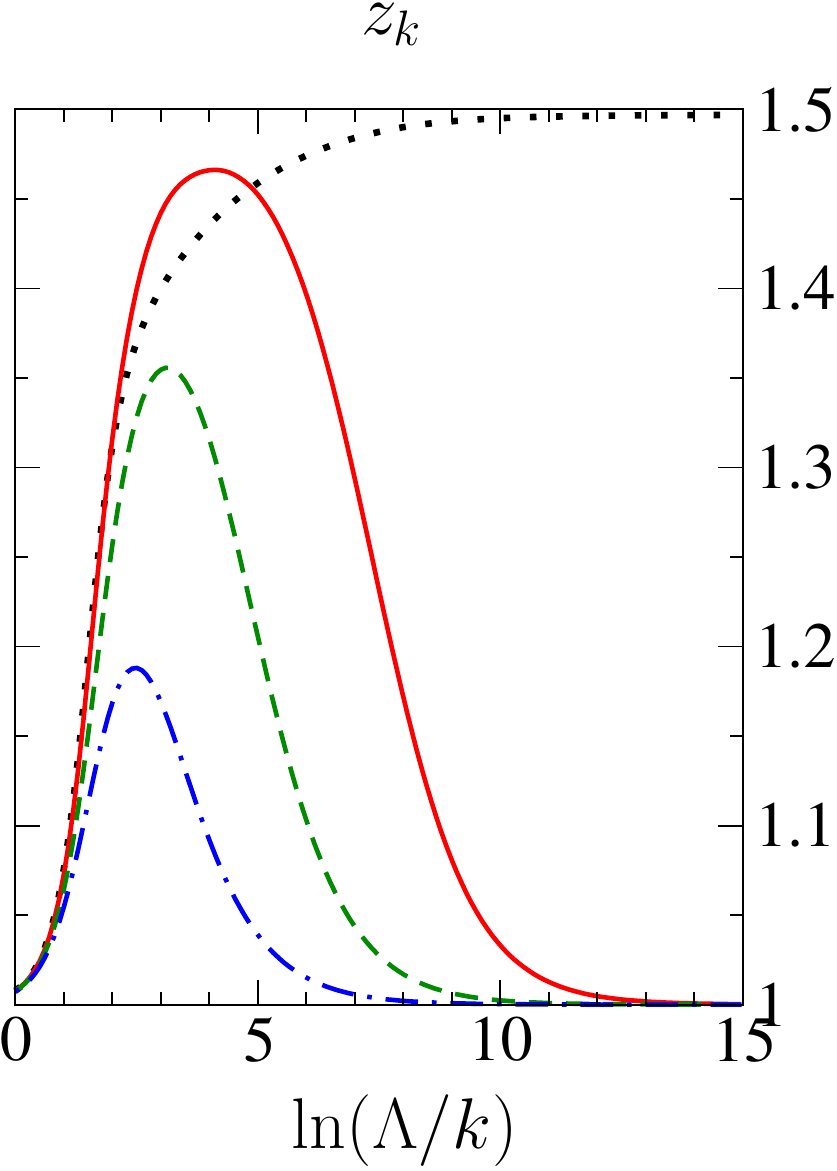}} 
\caption{(Left) Flow of the Luttinger parameter $K_k$ in the Bose-glass phase ($e^2=0$, dotted line) and in the presence of linearly confining interactions ($e^2/v\Lambda^2=0.1/0.4/0.8$) for $K_\Lambda=K=0.4$. (Right) Flow of the dynamical exponent $z_k$.} 
\label{fig_K} 
\vspace{0.25cm}
\centerline{\includegraphics[width=5.35cm]{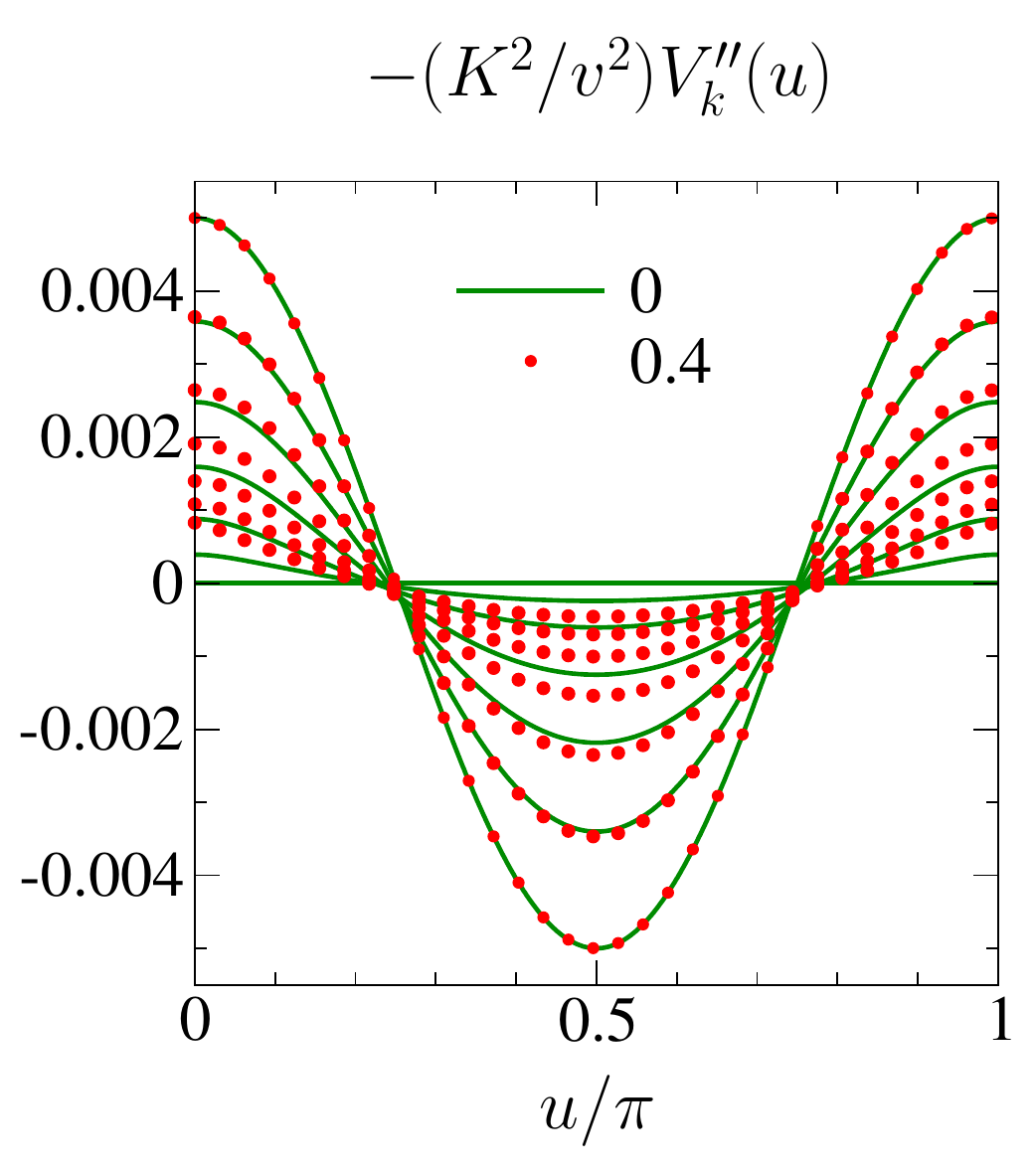}}
\caption{Potential $-(K^2/v^2)V''_k(u)=k^3 \delta_k(u)$ for $k$ varying between $\Lambda$ and $\Lambda\,e^{-15}$ (from top to bottom), $K=0.4$ and $8K^2\calD/v^2\Lambda^3=0.005$. Green lines: $e^2=0$ (Bose-glass phase), red circles: $e^2/v\Lambda^2=0.4$. } 
\label{fig_pot} 
\end{figure}

The linearly confining interaction dramatically alters this picture. After an initial decrease for $k\gtrsim k_x$, the Luttinger parameter $K_k$ stabilizes for $k\ll k_x$ to a nonzero value $K^*$ that depends on the initial conditions (i.e. the initial Luttinger parameter $K_\Lambda=K$, the strength of the disorder $\calD$, and the charge $e$). The running velocity $v_k$ takes the value $v^*=vK^*/K$ since $K_k/v_k=K/v$. The running dynamical exponent $z_k$ first increases (for $k\lesssim k_x$) but eventually converges to 1 for $k\to 0$ (Fig.~\ref{fig_K}). The potential $V_k(u)$ and its second derivative $V_k''(u)$ reach a nonzero value, while they vanish in the Bose-glass phase (in such a way that the dimensionless potential $\delta_k(u)$ reaches a nonzero fixed-value $\delta^*_{\rm BG}(u)$) (Fig.~\ref{fig_pot}). 

\begin{figure}
	\centerline{\includegraphics[width=3.88cm]{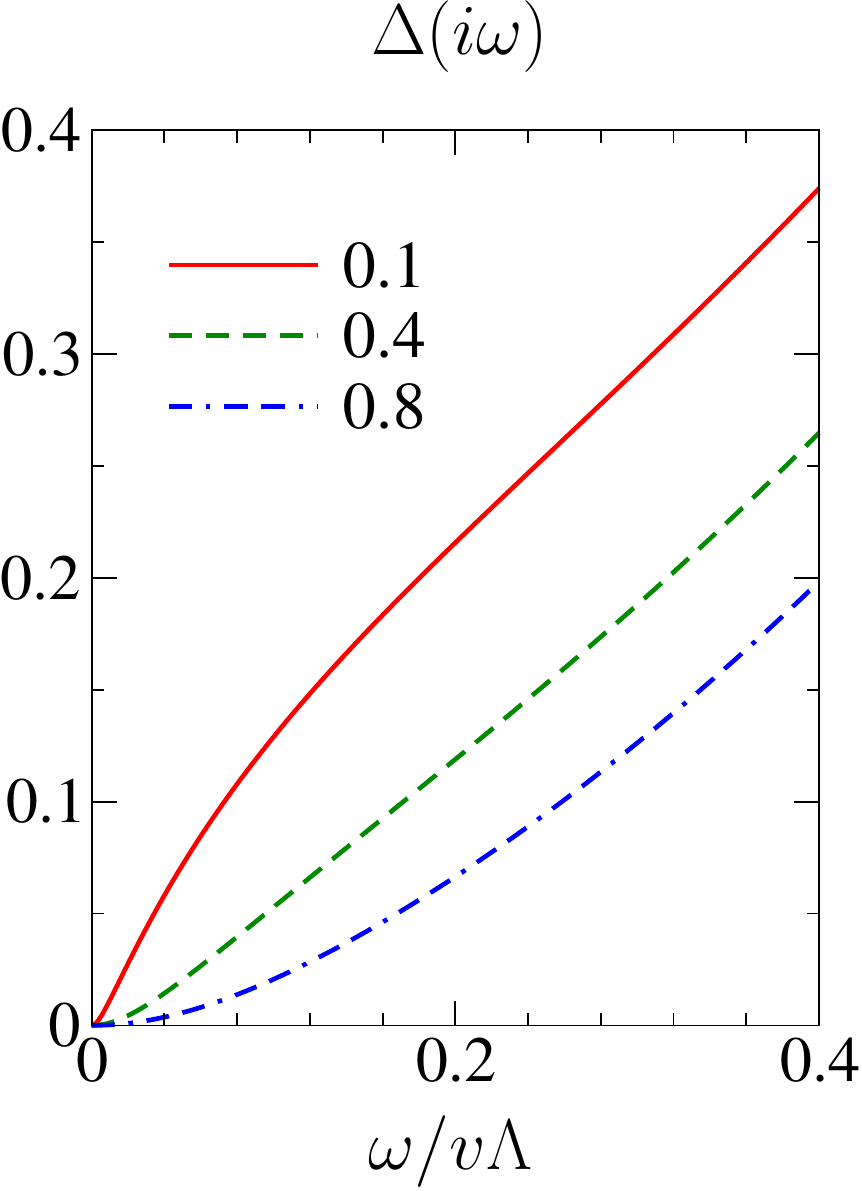}
		\includegraphics[width=4cm]{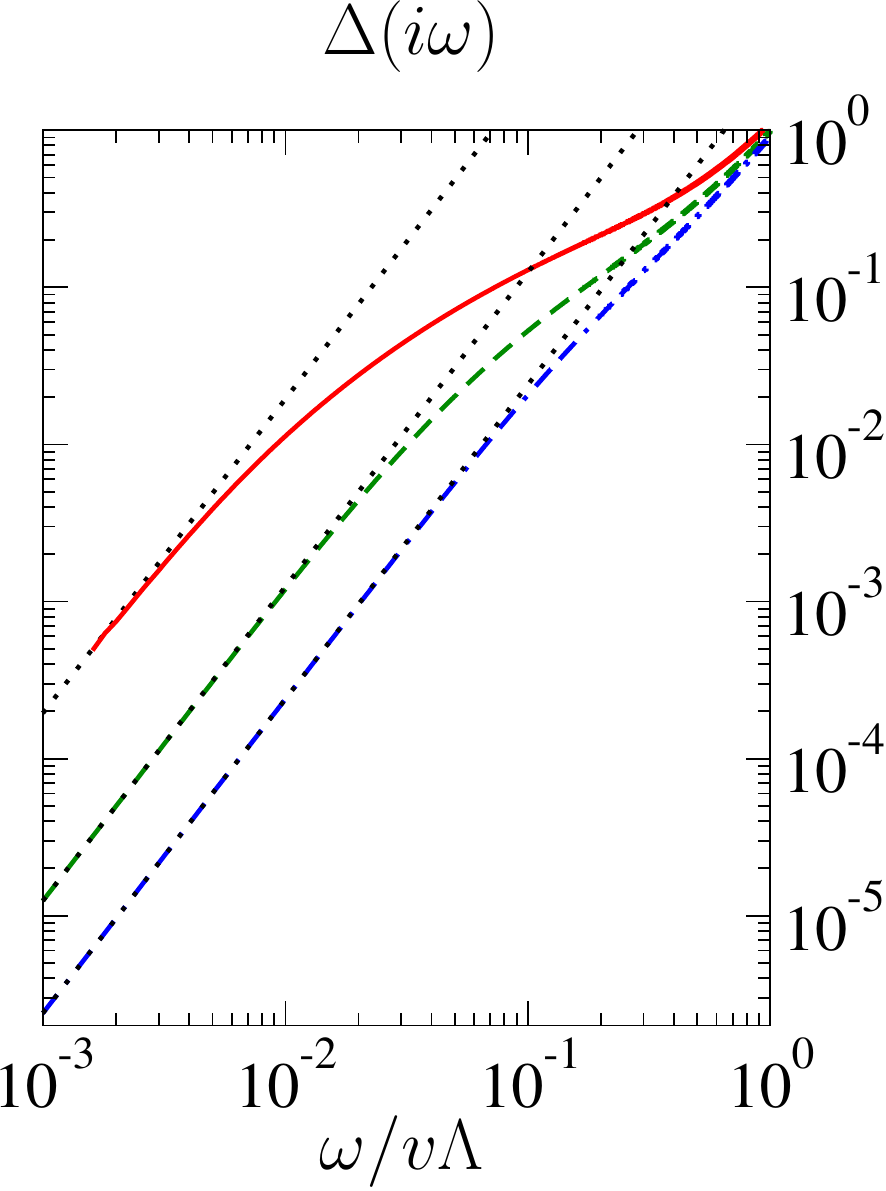}}
	\caption{(Left) Self-energy $\Delta(i\w)\equiv\Delta_{k=0}(i\w)$ vs $\w$ for $e^2/v\Lambda^2=0.1/0.4/0.8$. (Right) log-log plot showing the quadratic regime $\Delta(i\w)=Z_x\w^2/v^{*2}$ at low-frequencies (dotted lines) where the velocity $v^*=vK^*/K$ is deduced from $K^*=\lim_{k\to 0}K_k$ (see Fig.~\ref{fig_K}).
	The associated crossover frequency is $\w_x/v\Lambda=10^{-2.14}/10^{-1.24}/10^{-0.74}$ for $e^2/v\Lambda^2=0.1/0.4/0.8$ (from top to bottom).}
	\label{fig_SIG} 
\end{figure}

Figure~\ref{fig_SIG} shows the frequency dependence of the self-energy $\Delta(i\w)\equiv\Delta_{k=0}(i\w)$ (actually obtained for $k=\Lambda e^{-15}$). In the Bose-glass phase the self-energy $\Delta_k(i\w)$ is an analytic function of $\w^2$ for any nonzero value of $k$ (this is ensured by the presence of the infrared regulator term $\Delta S_k$ in the action and the fact that the cusp forms only at $k=0$), and behaves as $\Delta_k(i\w)\simeq Z_x\w^2/v_k^2$ for $|\w|\ll v_kk$. The low-frequency regime $\Delta_k(i\w)\sim \w^2$ is suppressed when $k\to 0$ and $\Delta(i\w)=A+B|\w|$ at low, but nonzero,  frequencies~\cite{Dupuis19,Dupuis20}. By contrast, in the Schwinger model, we expect the mass term $e^2/\pi^2$ in the propagator~(\ref{Pk}) to stop the RG flow at $k\sim k_x$ and therefore the self-energy $\Delta(i\w)$ to remain an analytic function of $\w^2$ in a finite frequency range, with the low-frequency behavior $\Delta(i\w)\simeq Z_x\w^2/v^{*2}$ for $|\w|\ll\w_x$ where $\w_x=v^*k_x=e\sqrt{K^*v^*/\pi}$. For small $e^2$, this quadratic regime is too narrow to be seen in the left panel of Fig.~\ref{fig_SIG} but is clearly apparent in a log-log plot. Furthermore, the approximation $\Delta(i\w)=Z_x\w^2/v^{*2}$ provides us with a very accurate fit of the numerical solution of the flow equation when $|\w|\ll\w_x$ (see the right panel of Fig.~\ref{fig_SIG}). Nevertheless, when the charge $e$ is sufficiently small, there is an intermediate frequency regime where $\Delta(i\w)\simeq A+B|\w|$ varies linearly as in the Bose-glass phase. 

The knowledge of the self-energy allows us to compute the density-density response function
\beq 
\chi_{\rho\rho}(q,i\w) = \frac{q^2}{\pi^2} \frac{1}{Z_x q^2 + \Delta(i\w) + e^2/\pi^2} . 
\eeq 
The vanishing of the compressibility when $e^2>0$, 
\beq
\kappa = \lim_{q\to 0} \chi_{\rho\rho}(q,i\w=0) = 0 , 
\eeq
which is due to the nonzero mass in $\chi_{\rho\rho}$, can be seen as a direct consequence of the STS. The conductivity is given by~\cite{Shankar90} 
\begin{align}
\sig(\w) &=  e^2 \lim_{q\to 0} \frac{-i\w}{q^2}\chi^R_{\rho\rho}(q,\w) \nonumber \\ 
&= \frac{-i\w e^2}{\pi^2[\Delta^R(\w)+e^2/\pi^2]} ,
\end{align}  
where the upper index $R$ denotes the retarded part of the correlation function obtained from the analytic continuation $i\w\to \w+i0^+$. Note that the expression of $\sig(\w)$ in terms of $\Delta^R(\w)$ is exact. Whether the conductivity exhibits a gap or not depends on the low-energy behavior of the self-energy. We have argued above that the self-energy $\Delta(i\w)$ is an analytic function of $\w^2$ below a strictly positive frequency threshold. Using the low-frequency limit $\Delta(i\w)=Z_x\w^2/v^*{}^2$, and therefore $\Delta^R(\w)=-Z_x\w^2/v^*{}^2$, one obtains
\beq 
\sig(\w) \simeq \frac{i\w\w_x^2}{(\w+i0^+)^2-\w_x^2 } 
\eeq 
for $|\w|\ll\w_x$. The conductivity is purely imaginary at low frequencies, which implies an optical gap.

\section{Discussion and conclusion}

The nonperturbative FRG approach, combined with bosonization, has proven to be successful in one dimension for the description of the Mott transition of bosons in a periodic potential (sine-Gordon model~\cite{Daviet19}) and the Bose-glass phase of disordered bosons~\cite{Dupuis19,Dupuis20}. In particular, it takes into account both solitonlike excitations and small fluctuations of the phase field $\varphi$ about its equilibrium value (this includes the soliton-antisoliton bound state (breather) of the sine-Gordon model) in sine-Gordon-like models. In this paper, we have shown that the FRG approach predicts the ground state of the disordered Schwinger model to exhibit a vanishing compressibility and a gapped optical conductivity.

This result is in striking disagreement with the GVM~\cite{Chou18}. The latter predicts the ground state of the disordered Schwinger model to be a Mott glass, i.e. a state intermediate between a Mott insulator and a Bose glass, characterized by a vanishing compressibility and a gapless optical conductivity. As compared to the FRG, the GVM appears to be a much less controlled approach; it leads to artifacts even in the case of the plain sine-Gordon model.\footnote{See, e.g., the discussion in Appendix C of Ref.~\cite{Giamarchi96}.}. An important limitation of the GVM is its inability to describe soliton and soliton-antisoliton pair excitations, which is yet crucial for a  proper analysis of the compressibility and optical conductivity in disordered one-dimensional quantum fluids~\cite{Nattermann07}. The GVM has nevertheless given an apparently satisfactory description of the Bose-glass phase~\cite{Giamarchi96}.\footnote{Whether small fluctuations of the phase (the only excitations that are considered in the GVM) are sufficient to explain the $\w^2$ dependence of the conductivity seems however controversial and contradicting results can be found in the literature; see, e.g.,
Refs.~\cite{Feigelman81,Aleiner94,Fogler02}.} This success however relies on the (somewhat ad hoc) choice of the marginal stability condition rather than the minimization of the free energy. The second choice gives a gapped optical conductivity and was discarded on physical grounds in Ref.~\cite{Giamarchi96}. The marginal stability condition was also chosen in the GVM approach to the disordered Schwinger model~\cite{Chou18}. Minimizing the free energy, which might give a gapped optical conductivity, could be a better choice and lead to results in agreement with the nonperturbative FRG.\footnote{Since, when minimizing the free energy, one finds a gapped conductivity for $e=0$, it is likely that the gap will persist in the case $e>0$.}  

Our results also disagree with the perturbative FRG which predicts a Mott glass in the disordered Schwinger model when the disorder is sufficiently strong~\cite{Giamarchi01}.  In the perturbative approach, the cusp in the disorder correlator $\delta_k(u)$ forms at a nonzero scale $k>0$ in the Bose-glass phase (i.e. for $e=0$). If the particle charge $e$ is sufficiently small, the cusp can still form before the renormalized (square) mass $\tilde e^2_k$ becomes of order unity and stops the flow; $\Delta^R(\w)$ then becomes nonanalytic near $\w=0$ and the conductivity is gapless. As discussed in detail in Ref.~\cite{Dupuis20}, the cusp formation at a nonzero momentum scale is an artifact of the perturbative approach, which treats in perturbation a diverging quantity ($\delta''_k(0)$); it does not survive in the nonperturbative FRG.

We therefore conclude that the existence of a Mott-glass phase in the disordered Schwinger model is very unlikely. 

\acknowledgments

I would like to thank Yang-Zhi Chou for entlighting discussions on the disordered Schwinger model and the work reported in Ref.~\cite{Chou18}. 


\end{document}